\documentclass[preprint,12pt]{elsarticle}

\usepackage{amssymb}
\usepackage{amsmath}

\usepackage{multirow}

\journal{Optics Communications}

\begin{document}

\begin{frontmatter}



%\title{Release-recapture measurements with cold $^{85}$Rb using optical nanofibres}
\title{Measurements on release-recapture of cold $^{85}$Rb  atoms using an optical nanofibre in a magneto-optical trap}


\author[ucc,tyn,oist]{L.~Russell}
\ead{laura.russell@oist.jp}

\author[ucc,tyn,oist]{R.~Kumar}

\author[ucc,tyn,cat]{V. B.~Tiwari}

\author[ucc,oist,ukzn]{S.~Nic Chormaic}

\address[ucc]{Physics Department, University College Cork, Cork, Ireland}
\address[tyn]{Tyndall National Institute, Lee Maltings, Propsect Row, Cork, Ireland}
\address[oist]{Light-Matter Interactions Unit, OIST Graduate University, 1919-1 Tancha, Onna-son, Okinawa 904-0495, Japan}
\address[cat]{Laser Physics Applications Section, Raja Ramanna Centre for
Advanced Technology, Indore 452013, India}
\address[ukzn]{School of Chemistry and Physics, University of Kwa-Zulu Natal, Durban 4001, South Africa}

\begin{abstract}
We have performed release-recapture temperature measurements of laser-cooled $^{85}$Rb atoms using an optical nanofibre (ONF) in a magneto-optical trap (MOT). The effects of changing the cooling laser light-shift parameter on the temperature of the cold atoms and spring constant of the trap are studied. By varying the cold atom number density near the ONF, the onset of the multiple scattering regime is observed without the need for an estimation of the atom cloud size. Moreover, this sensitive ONF assisted release-recapture technique is easily able to detect any optical misalignment of the cooling laser beams in the MOT.
\end{abstract}


\end{frontmatter}

% \linenumbers

\section{Introduction}
\label{Introduction}
Temperature is undeniably one of the most important characteristics of a laser-cooled sample of atoms that one can measure. By experimentally determining the ensemble temperature, $T$, quantities such as spring constant and diffusion coefficient can be estimated \cite{Kohns1993,Wallace1994}, temperature and density regimes can be mapped out \cite{Vorozcovs2005}, and a wealth of information regarding the efficiency of the trapping, cooling and compression scheme can be revealed \cite{Steane1991,Tung2000}. Although there now exist many temperature measurement techniques, for example \cite{Meacher1994,Silva2006}, the most commonly-implemented methods use the thermal expansion of the cloud --  time-of-flight (TOF) measurements -- to estimate $T$ \cite{Chu1985,Lett1988}. In this paper, we focus on the release-recapture (RR) which is sensitive to the velocity distribution of the cloud. For temperatures at the Doppler limit (144 $\mu$k for $^{85}$Rb) and above, as is the case in this work, the RR method is well-suited. At temperatures siginificantly below the Doppler limit, gravity begins to play a role in the thermal expansion of a cold cloud of atoms because the initial velocity of the atoms becomes small compared to the velocity acquired due to gravity during the expansion phase.

An optical nanofibre (ONF) \cite{Love1991,Brambilla2010} is generally made from standard optical fibre which is heated and simultaneously pulled to produce a subwavelength diameter fibre. ONFs can be used to couple light into optical resonators \cite{Knight1997,Gorodetsky1999,Ward2010,Watkins2012a} and for characterising and guiding particles \cite{Brambilla2007,Frawley2012a}. Spontaneous emission into the guided modes of an optical nanofiber is enhanced \cite{LeKien2005}, making it an ideal high-sensitivity tool for channeling atomic fluorescence to a detector. Thus, in recent years, ONFs have also been shown to act as a measurement tool and delivery platform for cold atoms \cite{Nayak2007,Russell2009,Minogin2009,Morrissey2009,Vetsch2010,Russell2012,Russell2012a,Goban2012} and atomic vapours \cite{Hendrickson2010,Spillane2008}.

Here, the RR method is performed by positioning a cold cloud of $^{85}$Rb atoms centrally around an ONF. Previous works show that, despite the presence of the hot ONF surface in the cloud of cold atoms, sub-Doppler temperatures can still be obtained with large red-detunings of the cooling laser beams \cite{Russell2012a}.

\section{Background}
\label{Background}
If a spherical cloud of atoms, with a Gaussian velocity distribution, is allowed to expand homogeneously from an initial finite diameter, the fraction of atoms, $f_r$, remaining after the release time, $\Delta t_2$, is given by

\begin{equation}
f_r=\frac{1}{\pi^{3/2}}\int_0^{v_c/v_T} e^{-u^2}u^2du.4\pi,
\label{eqn:equation1}
\end{equation}

where $ u^2du.4\pi $ is the spherical polar coordinate for velocity. The thermal velocity of the atoms in the MOT at a temperature $T$ is $v_T=\sqrt{2k_BT/m}$ and $v_c=R_c/\Delta t_2$ is the velocity at which the atoms just reach a position $R_c$ in the time interval $\Delta t_2$. The capture region is characterised by the radius of the MOT beams, $R_c$. Integrating Equation \ref{eqn:equation1} yields:

\begin{equation}
f_r=-\frac{2 e^{-\frac{v_c^2}{v_T^2}} v_c}{\sqrt{\pi } v_T}+\text{Erf}\left[\frac{v_c}{v_T}\right].
\label{eqn:equation2}
\end{equation}

Equation \ref{eqn:equation2} describes $f_r$ as a function of $\Delta t_2$. This equation is fitted to the experimental data by taking $R_c$ as known and $v_T$ is the fitting parameter.

\section{Experiment}
\label{Experiment}
The cold $^{85}$Rb atom cloud is formed with a standard MOT configuration of three orthogonal, counter-propagating cooling beams intersecting at the centre of an inhomogeneous magnetic field. The cooling laser is locked to the crossover peak, $^2S_{1/2}F_g=3\rightarrow$ $^2P_{3/2}F_e=(2,4)_{co}$ and then red-detuned from the cooling transition using an acousto-optical modulator (AOM). This detuning is controlled by the frequency input (0--10 V) to the AOM driver. Each beam has a maximum diameter of 24 mm (controlled via an aperture). The repumping laser is locked to the crossover peak, $^2S_{1/2}F_g=2 \rightarrow$$^2P_{3/2}F_e=(1,3)_{co}$. The magnetic field is created by a pair of coils, each carrying currents of  $\sim$4 A in opposite directions, to generate an axial field gradient in the region of 15 G/cm at the centre of the MOT. Between 10$^4$ and 10$^8$ atoms are trapped in the MOT depending on experimental parameters.

To fabricate the ONF, the usual heat-and-pull-technique \cite{Ward2006} is used with Fibercore SM750 fibre. For these experiments, the transmission of the ONF was $\sim$60\% and the waist diameter was $\sim 1\mu$m (as determined with a scanning electron microscope). Further details about the experimental characteristics of an ONF can be found in \cite{Morrissey2009}. The fibre pigtails are coupled into and out of the UHV chamber using a Teflon feed-through \cite{Abraham1998} and one end is connected to a single photon counting module (SPCM\footnote{Perkin and Elmer single photon counting module; model: SPCM-AQR-14; dark count = 100 counts $s^{-1}$; quantum efficiency = 60\% at 780 nm}). The cloud is aligned centrally around the ONF using imaging techniques and optimisation of the fluorescence coupling into the fibre guided modes is detected by the SPCM. Final adjustments to cloud position are made with magnetic coil currents.

\subsection{Procedure}
\begin{figure}[!h]
  \centering
   \includegraphics[width=12cm]{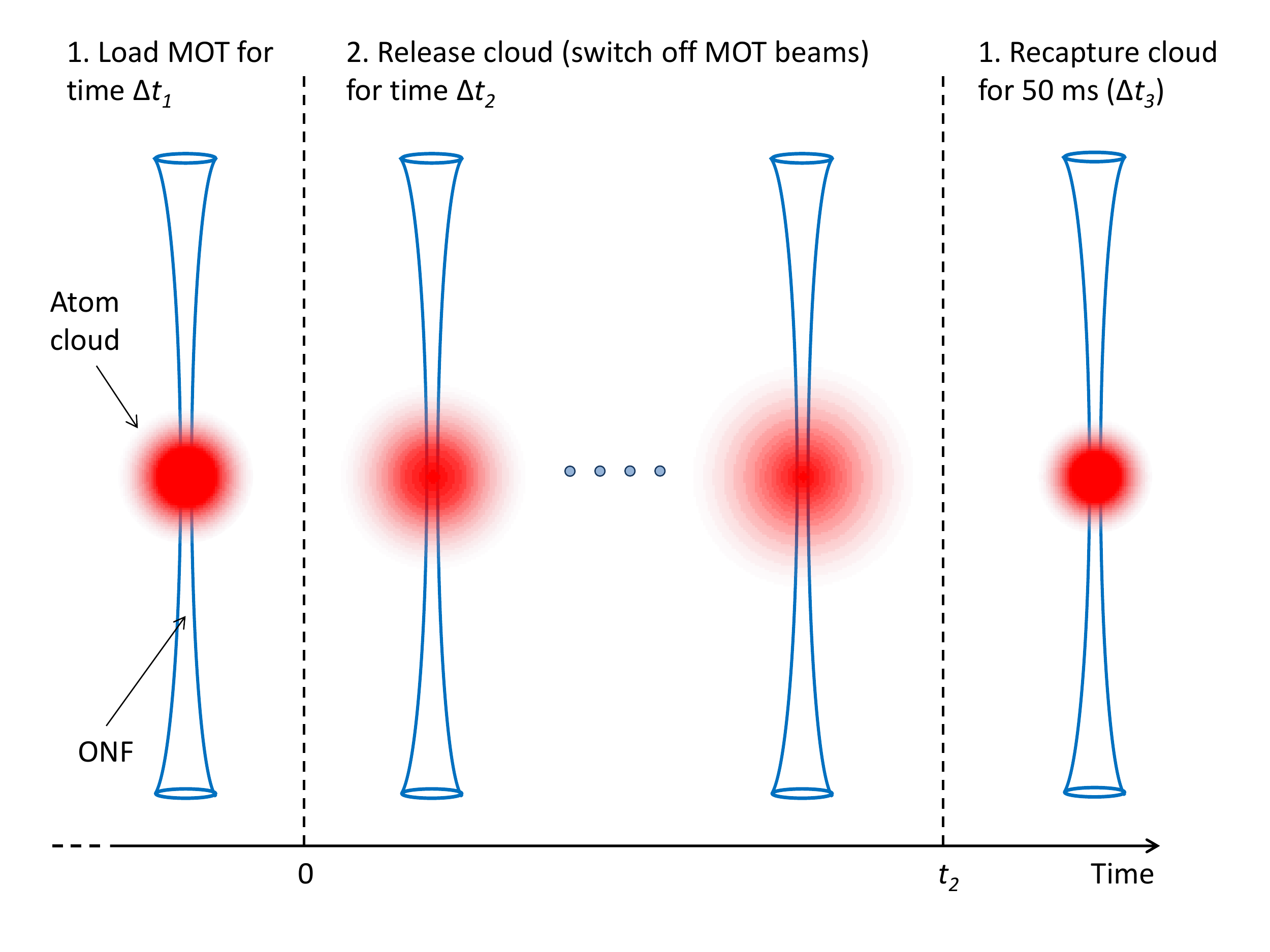}
  \caption{Illustration of the release and recapture sequence as a function of time. The cloud is positioned centrally around the ONF. The cloud is loaded for a time $\Delta t_1$ and then released from the trap for a time $\Delta t_2$. $\Delta t_2$ is varied from sequence to sequence to build up a profile of the velocity of the atoms in the MOT. The recapture time, $\Delta t_3$, is set to 50 ms. The cloud is then released once more for 50 ms, before repeating the entire sequence.}
   \label{fig:figure1}
\end{figure}

The cooling laser beams were switched on and off with an AOM in order to achieve the desired loading, release and recapture sequence (see Figure \ref{fig:figure1}). A recapture time, $\Delta t_3$, of 50 ms was used to ensure that no background vapour atoms are recaptured by the MOT. The loading time of the MOT is typically $\sim$1 second. The magnetic field stays on at all times.

\begin{figure}[!h]
  \centering
   \includegraphics[width=12cm]{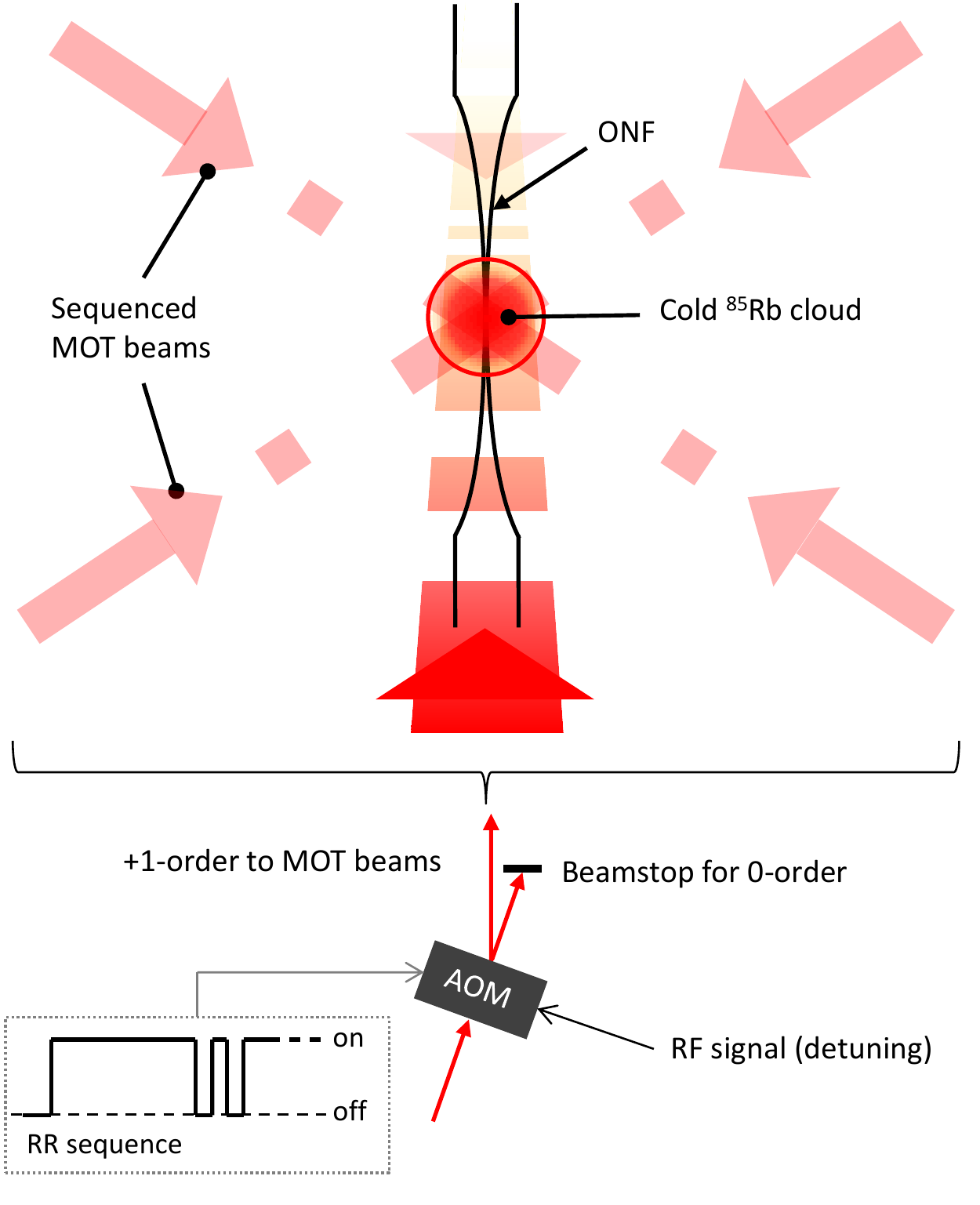}
  \caption{Illustration of the experimental setup. The 1$^{\text{st}}$-order beam from an AOM is expanded and split into three equal-intensity cooling beams for the MOT. The $^{85}$Rb cloud is positioned centrally around the ONF. The cooling beams are switched on and off according to the sequence shown here with an AOM. Beam detuning is controlled via the tunable RF input on the AOM.}
   \label{fig:figure2}
\end{figure}

To perform a temperature measurement, the cloud of atoms was loaded for 10 seconds ($\Delta t_1$) to ensure a steady-state number of atoms was reached. Then, the cooling laser was switched off using the AOM for a time $\Delta t_2$ to allow the cloud to expand freely. $\Delta t_2$ is varied each time the sequence is repeated: $\Delta t_2$ = 5 ms, 10 ms, 20 ms, ... 150 ms. After the release time has passed, the cooling laser is switched on for $\Delta t_3$=50 ms to recapture the cloud. The cooling beams are switched off again with the AOM for 50 ms to provide a suitable contrast between the signal from the recaptured atoms and the background level. The sequence is recommenced for the next value of $\Delta t_2$. An image of the cloud is recorded using a CMOS camera and image analysis is performed to estimate the cloud radius.

Following this release and recapture process, fast atoms escape the MOT after a short release time and slower atoms are lost only after longer release times. To estimate the temperature of the atoms, the fraction of remaining atoms is calculated as a function of $\Delta t_2$. This is proportional to the fluorescence coupled into the nanofibre. This method is sensitive to the velocity distribution of the cloud.

\section{Results and discussion}
\begin{figure}[!htbp]
  \centering
   \includegraphics[width=13cm]{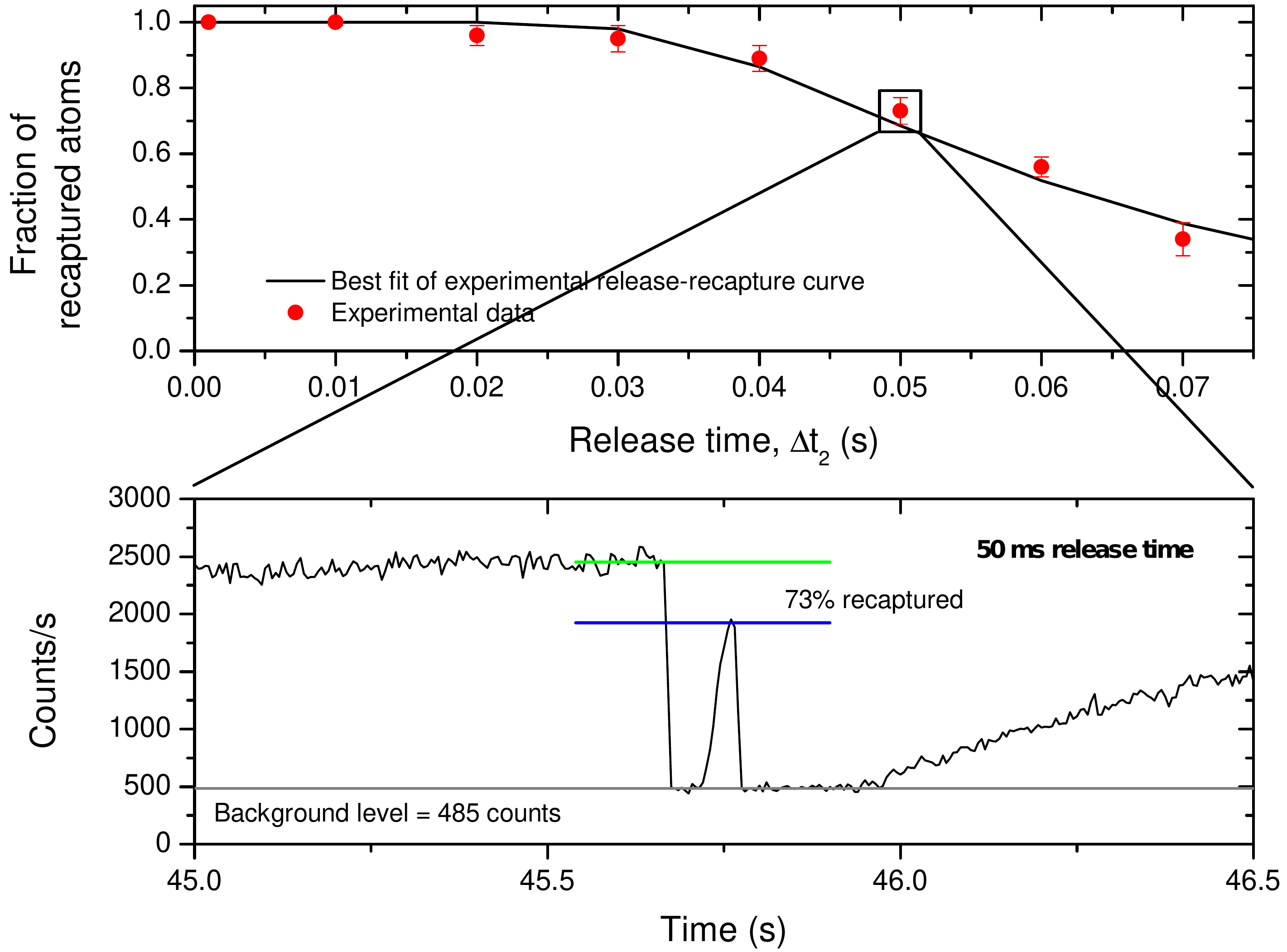}
 \caption{Fraction of recaptured atoms estimated from the fluorescence data obtained from an optical nanofibre placed near the cold cloud of atoms. For a release time, $\Delta t_2$, of 50 ms, 73\% of the expanded cloud was recaptured with $\delta$=-2.6$\Gamma$ and cooling laser intensity of 2.9$I_s$ (4.6 mW/cm$^2$) per beam.
%(Data from '13-08-12' for 6.4 V detuning)
}
   \label{fig:figure3}
\end{figure}

Temperatures approaching the Doppler limit have been observed at moderate detunings when the alignment of MOT beams has been particularly good. For example, for a detuning of -2.6$\Gamma$ (where $\Gamma=2\pi \times 5.9$ MHz is the natural linewidth of the $5^2S_{1/2}\rightarrow5^2P_{3/2}$ transition in $^{85}$Rb) with a cooling laser intensity per beam, $I_{beam}$, of 2.9$I_s$ (where $I_s$=1.6 mW/cm$^2$ for $\sigma ^{\pm}$-polarised light on the $^{85}$Rb cooling transition), $T$ is estimated to be 167 $\mu$K (Figure \ref{fig:figure3}). Poor MOT beam alignment would mean that atoms may leave the capture region in a non-isotropic way, the signature of which is an immediate and sharp decrease in recaptured atoms. The importance of precise optical alignment and power equalisation in the MOT beams is well known \cite{Straten2002,Townsend1996}. For example, \cite{Straten2002} reports a temperature and associated variation of \mbox{(147$\pm$25) $\mu K$} depending on the alignment of the laser beams. By using the ONF as the detection tool, the effect of beam misalignment and power mismatching is detectable with a greater sensitivity than for fluorescence imaging techniques.

\begin{figure}[!htbp]
  \caption{Temperature as a function of cooling laser intensity normalised to the saturation intensity, $I_s$ ($\delta$=-2$\Gamma$). The solid red line is a linear fit to the experimental data.}
  \centering
   \includegraphics[width=12cm]{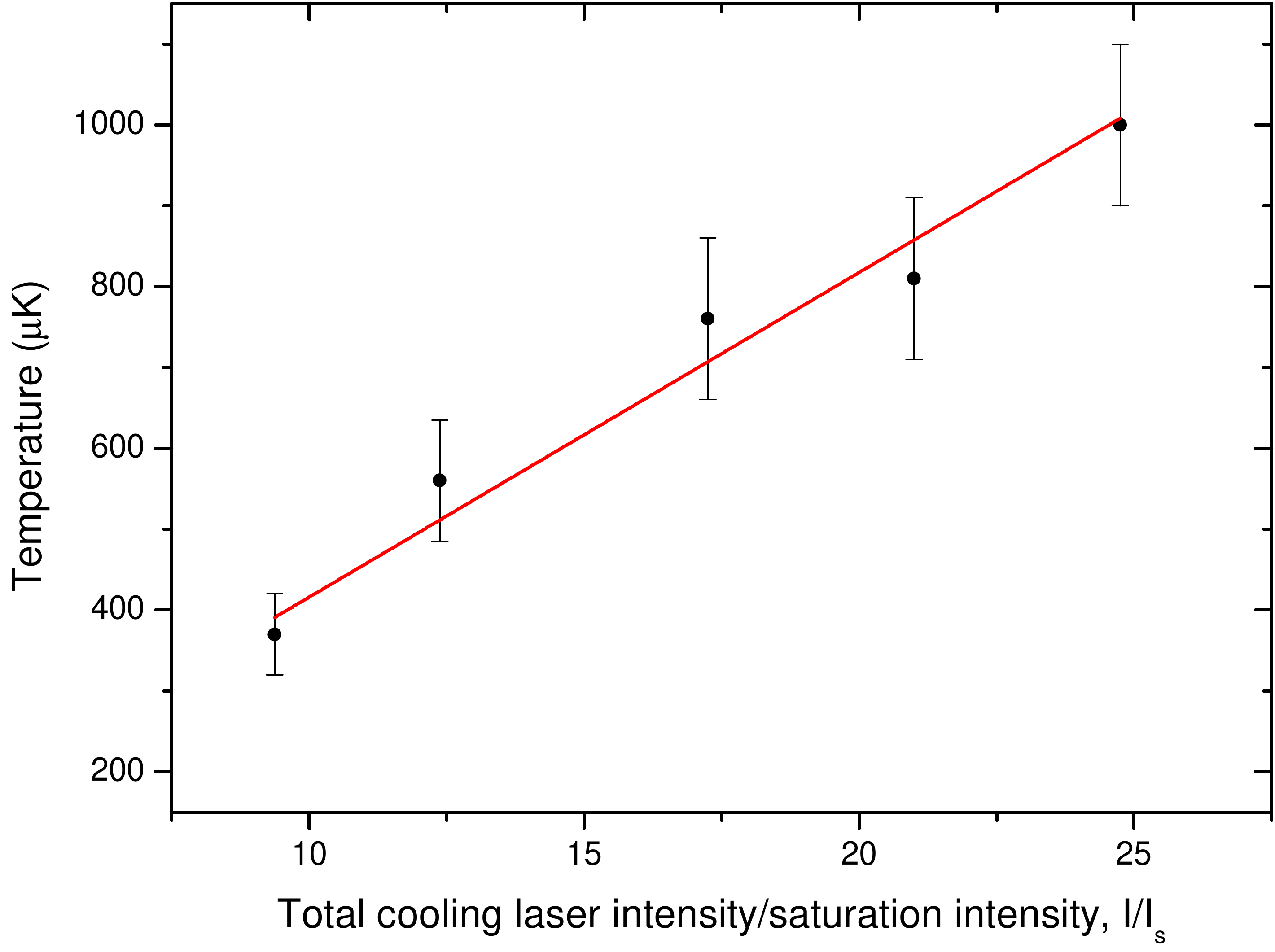}
   \label{fig:figure4}
\end{figure}

The temperature of the cold atoms as a function of $I_{beam}$ was investigated (Figure \ref{fig:figure4}) and measurements show that a span of a few hundred $\mu$K can be observed when varying $I_{beam}$ from 2.5 mW/cm$^2$ to 6.7 mW/cm$^2$ at a constant cooling laser red-detuning of $\sim$2$\Gamma$. As expected, the temperature reduces as laser intensity is reduced \cite{Lett1989a}.

\begin{figure}[!ht]
  \centering
   \includegraphics[width=12cm]{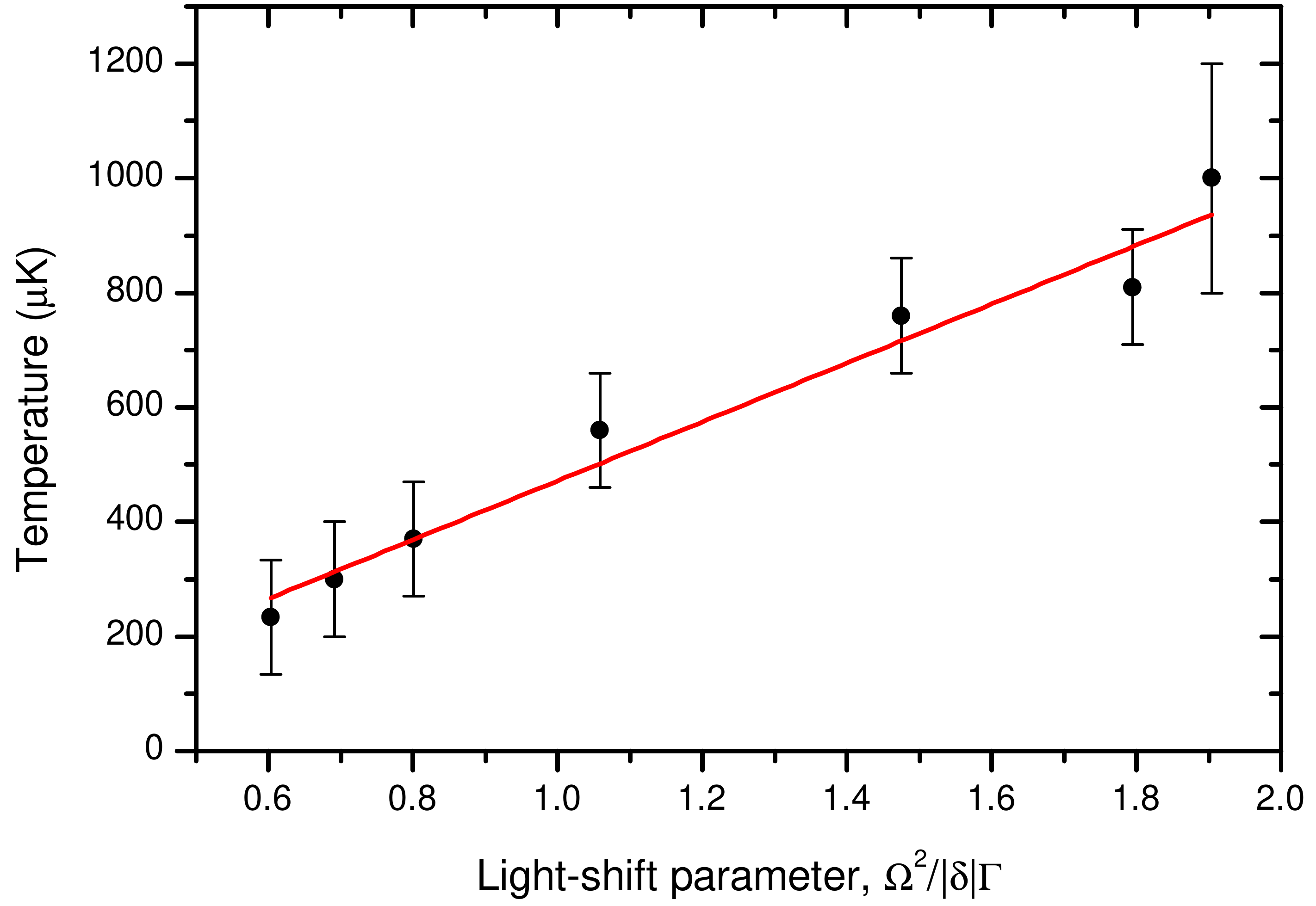}
  \caption{Temperature of the cloud as measured with RR plotted against the dimensionless light-shift parameter $\Omega ^2/|\delta|\Gamma$ (black circles) and a linear fit (red line) to the data.}
   \label{fig:figure5}
\end{figure}

\begin{figure}[!ht]
  \centering
   \includegraphics[width=12cm]{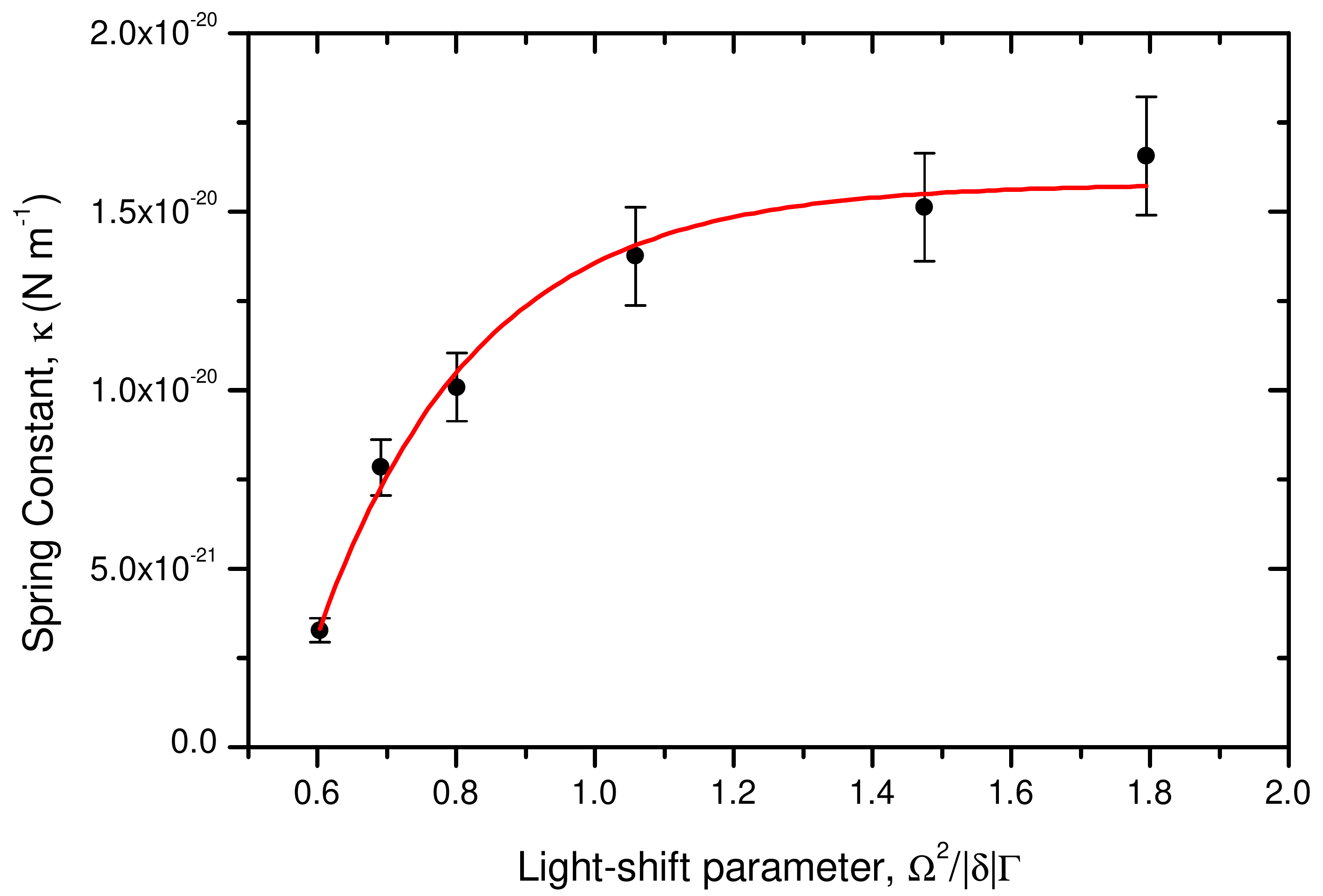}
  \caption{Spring constant against dimensionless light-shift parameter $\Omega ^2/|\delta|\Gamma$ (black circles) with a fit for a guide to the eye (red, solid line).}
   \label{fig:figure6}
\end{figure}

Traditionally, $R_c$ is the quantity with the greatest uncertainty. A small change in $R_c$ ($\sim$ 2.5 mm) when fitting Equation \ref{eqn:equation2} to the data results in a temperature shift of the order of the error bars seen in Figures \ref{fig:figure4}--\ref{fig:figure6}.

In order to compare the RR temperature measurements with another technique, we have taken data directly from previous work which used the method of forced oscillations to estimate cloud temperature (see \cite{Russell2012a} for full details). Table \ref{tab:temperatureVdetuning} presents a comparison of results for temperature as a function of cooling laser red-detuning. The results found using each technique correlate quite well. In particular, for higher detunings, the $T$ values which have been estimated with each method agree more strongly. Lower values of red-detuning were not easily examined with the ONF used for RR as the sacrifice in $N_A$ was sufficient to lower the fluorescence coupling signal siginificantly.

\begin{table}[!htbp]
\centering
\begin{tabular}{lccc}
\hline
\noalign{\smallskip}
\multirow{2}{*}{\textbf{Method}} & Red Detuning & Temperature \\
 & $\Gamma$ & mK \\
\hline
\noalign{\smallskip}
\multirow{4}{*}{\emph{Release recapture}} & 1.1 & 1.80$\pm$0.20 \\
 & 2.0 & 0.56$\pm$0.13 \\
 & 3.5 & 0.23$\pm$0.03 \\
\hline
\noalign{\smallskip}
\multirow{5}{*}{\emph{Forced oscillation}} & 2.1 & 0.97$\pm$0.08 \\
 & 2.6 & 0.65$\pm$0.06 \\
 & 3.1& 0.37$\pm$0.04 \\
 & 3.5 & 0.18$\pm$0.02 \\
\end{tabular}
\caption{Variation of temperature with cooling laser red detuning in units of the natural linewidth using two different measurement techniques. I$_{beam}$ was approximately 2.2$I_s$ for release-recapture and 1.3$I_s$ for forced oscillation. Forced oscillation data is taken directly from \cite{Russell2012a}.}
\label{tab:temperatureVdetuning}
\end{table}

From Figure \ref{fig:figure5}, it is clear that temperatures obtained with the RR technique increase linearly with the light-shift parameter, $\Omega^2/|\delta|\Gamma$, as expected \cite{Wallace1994}. The spring constant, $\kappa$, can be inferred from these temperature values.  $\kappa$ describes the restoring force in the MOT and is a particularly relevant quantity in relation to compressing atoms to high density. To determine $\kappa$ it is assumed that, in thermal equilibrium, the atom cloud has a thermal energy given by $k_BT=\kappa \langle r^2 \rangle=m\langle v^2 \rangle$, where $T$ is the experimentally-determined cold atom cloud temperature, $k_B$ is Boltzmann's constant, $r$ is the radius of the atom cloud, and $\langle v^2 \rangle $ is the mean square atomic velocity \cite{Kohns1993,Steane1991}. For each value of red-detuning, the cloud radius is estimated using image analysis. Figure \ref{fig:figure6} shows that the spring constant increases with intensity at low intensity or detuning values and then levels off at some critical value of the light-shift parameter. Wallace \emph{et al}. \cite{Wallace1994} report that the spring constant is not independent of $I_{beam}$ until a moderately high value of the light-shift parameter is reached. It is interesting to observe that, as evident in Figure \ref{fig:figure6}, $\kappa$ levels off above light-shift parameter values of $\approx$1. As the most dense part of the cloud is being studied with the ONF, this may be an interesting observation when compared to measurements done with photodiodes.

As atom number is increased in the MOT, the regime of operation transitions from temperature-limited (TL) to multiple-scattering (MS) \cite{Walker1990,Townsend1995,Tiwari2008}. In the TL regime, the cold atoms are essentially non-interacting because $N_A$ is small (typically less than $10^4$) and the density is low. The cloud acts as an ideal gas in this regime, and, as more atoms are loaded into the trap, the size of the cloud remains the same while the density increases linearly and its distribution remains Gaussian \cite{Guedes1994}. For larger $N_A$ ($\ge 10^5$) the cloud begins operating in the MS regime and the density becomes largely independent of $N_A$ \cite{Sesko1991a}. Two effects are seen in this regime. Firstly, repulsive forces caused by the re-absorption of scattered photons lead to an increase in the cloud size while the density remains constant. This radiation trapping effect determines the density and temperature of cold atoms \cite{Vorozcovs2005,Overstreet2005}. For example, at $N_A\sim 10^7$ the cloud density is maintained and the optical thickness is such that, on average, each photon absorbed from the cooling laser beams will, after re-emission, scatter no more than once on its way out of the cloud. This determines the spatial growth of the cloud as $N_A$ increases beyond $10^7$ \cite{Lindquist1992}. The second effect is due to an attenuation force which arises from the intensity gradients in the MOT beams and the absorption of such by the atoms \cite{Sesko1991a,Dalibard1988}. This results in spatial compression of the cloud and a small reduction in spring constant of the trap \cite{Greenberg2007}. When the MOT transitions from the TL to the MS regime, the atom distribution may or may not change from Gaussian to flat-topped. Thus, at higher cloud densities, cloud images may not be a direct indicator of density. Measurements using an ONF negate the use of imaging analysis to estimate cloud volume (for example, \cite{Overstreet2005}) making it simpler to observe regime-change in the MOT via fluorescence coupling.

By considering an observation volume surrounding the ONF, cloud density can be studied as the light-shift parameter is varied. It is assumed that atoms within a hollow observation cylinder with an outer radius equal to the ONF radius + 300 nm are most likely to emit fluorescence into the guided mode of the ONF \cite{Nayak2007,Morrissey2009}. This number of effective atoms, $n_{eff}$, can be estimated at one end of the ONF using $n_{eff}=2C_P/R_{sc}\eta_{ONF}Q\eta_{QD}$ where $R_{sc}$ is the atomic scattering rate, $\eta_{ONF}$ is the average coupling efficiency of photons into the guided ONF mode in one direction (estimated at 2\% using previous work based on $^{133}$Cs \cite{LeKien2006}), $Q$ is the ONF transmission from the middle of the ONF waist to the detector (the transmission through the entire length of fibre is 60\% so $Q$=77\% for half the fibre length), and $\eta_{QD}$ is the quantum efficiency of the SPCM (60\%). The quantity $C_P$ is the fluorescent count rate obtained by the SPCM and is obtained from the RR raw data. The scattering rate, $R_{sc}$ is described by \cite{Townsend1995}

\begin{equation}
R_{sc}=\frac{\Gamma}{2}\frac{C_1^2\Omega _{tot}^2/2}{\delta ^2+\Gamma ^2/4+C_2^2\Omega _{tot}^2/2}.
\end{equation}

Here, $\Omega _{tot}$ is the Rabi frequency for all MOT beams and is taken to be six times that of any one of the trapping beams. $C_1$ and $C_2$ are average Clebsch-Gordan co-efficients. The values of $C_1^2$ and $C_2^2$ are assumed to be equal due to optical pumping among the Zeeman sublevels in the presence of strong coupling between atoms and the radiation field \cite{Townsend1995}.

\begin{figure}[!ht]
  \centering
   \includegraphics[width=14cm]{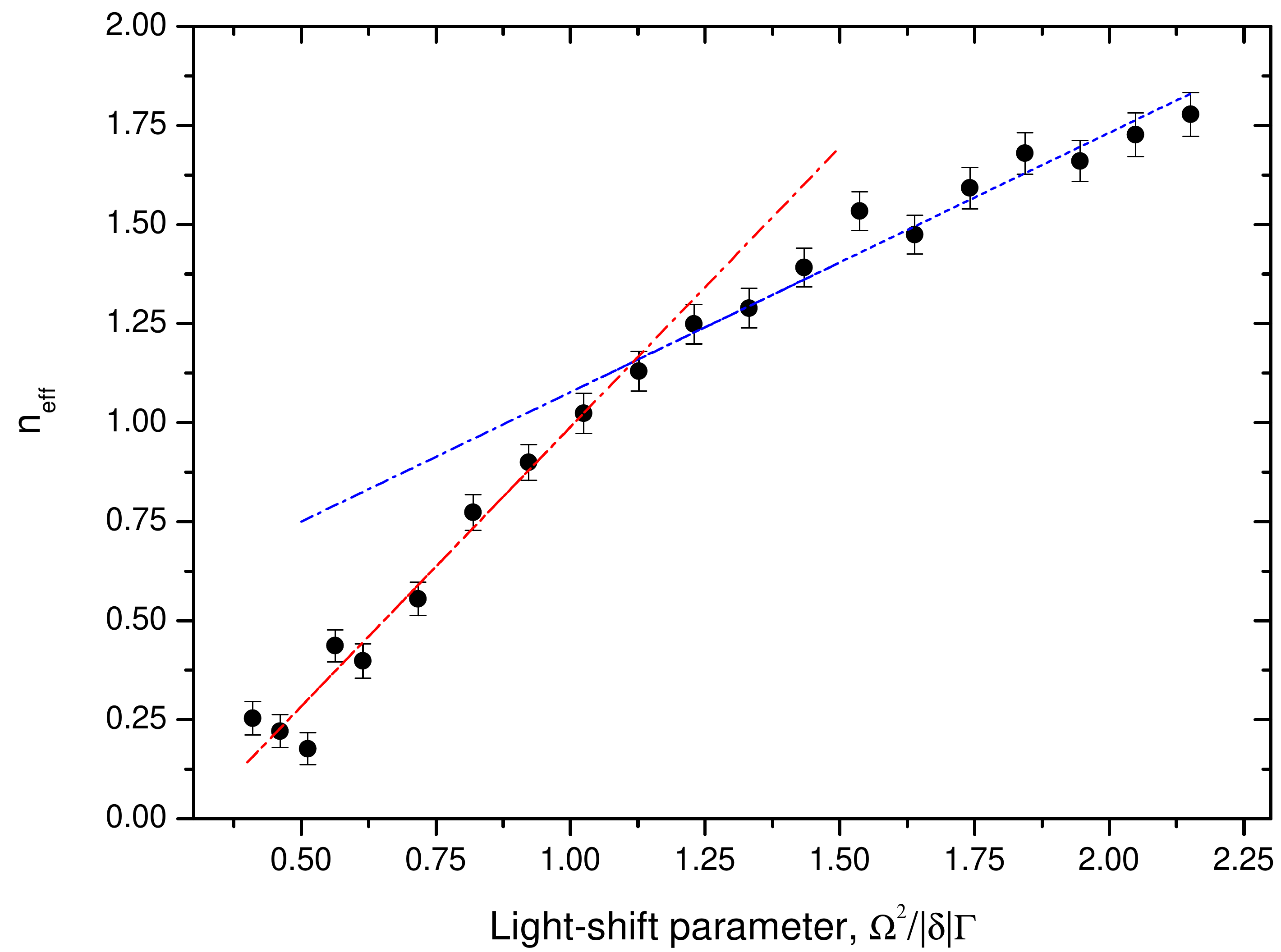}
  \caption{Estimation of effective atom-number, $n_{eff}$, with increasing dimensionless light-shift parameter $\Omega ^2/|\delta|\Gamma$ (black circles) with two linear fits to indicate the change in slope around a light-shift parameter 1.2.}
   \label{fig:figure7}
\end{figure}

If $n_{eff}$ is plotted as a function of light-shift parameter, saturation of the atom number commences from $\Omega^2/|\delta|\Gamma \sim$ 1.2 onwards (Figure \ref{fig:figure7}). As the observation volume is fixed by the ONF radius, this saturation effect is due to the spatial expansion of the cloud while it maintains constant density and may also indicate the onset of the MS regime in the MOT.

\section{Outlook and conclusion}
The temperature of a cold ensemble of $^{85}$Rb atoms has been measured using the RR method with an ONF. The RR sequence was applied to an AOM allowing the MOT beams to be switched off rapidly. The results presented here agree with those found by the forced oscillation method \cite{Russell2012a} and again reinforce the viability of placing the ONF in yet colder atomic samples. This work has highlighted the sensitivity of the cloud temperature to small changes in beam misalignment, detunings, and intensities, and demonstrates the ability to detect these temperature changes with the ONF. As the detector can be positioned anywhere in the cloud of atoms, a systematic temperature measurement, while exploring the entire parameter space, can be performed.

Free-space RR measurements show that, with the same variation in laser cooling intensity as we have used, a few hundred $\mu$K span can be observed. The results here display this trend and, additionally, provide detailed information about the velocity distribution of the atom cloud. In particular, with the ONF-based RR method, the system is sensitive to atoms near the fibre surface and, using data analysis, it is possible to generate a decay curve that shows how those near-surface atoms with a high velocity leave the capture region quickly and do not continue to contribute to the signal detected via the ONF.

Values for the springs constant of the MOT, inferred from temperature results, increase with cooling laser intensity until some critical value of the light-shift parameter is reached ($\approx$1) and then begin to level out. Further, by examining coupling signal strengths while varying the light-shift parameter, the density regime in which the MOT is operating can be identified.

\section{Acknowledgements}
The authors wish to thank W. Cotter for help in creating the AOM sequencing programme. This work is partially supported by Science Foundation Ireland under grant no.s 07/RFP/PHYF518 and 08/ERA/I1761 through the NanoSci-E+ Project NOIs, OIST Graduate University and the Higher Education Authority via the INSPIRE programme. LR acknowledges support from IRCSET under the Embark Initiative.



%% References


\bibliographystyle{model1-num-names}
\bibliography{refs}






% \end{thebibliography}

\end{document}